\documentclass[prl,twocolumn,amsmath,amssymb]{revtex4-1} %

\usepackage{epsfig,amsmath}
\usepackage{subfigure}
\usepackage{graphicx}
\usepackage{dcolumn}
\usepackage{stmaryrd}
\usepackage{mathrsfs}
\usepackage{pifont}
\usepackage{amsthm}
\usepackage{amssymb}
\usepackage{bm}
\usepackage{latexsym}
\usepackage[colorlinks=true,linkcolor=blue,citecolor=blue]{hyperref}
\usepackage{color}

\theoremstyle{plain}

\def\oe#1{{ Opt.\ Express} {\bf#1}}

\def\pra#1{{ Phys.\ Rev. A\/} {\bf#1}}

\def\pre#1{{ Phys.\ Rev. E\/} {\bf#1}}
\def\prl#1{{ Phys.\ Rev.\ Lett.} {\bf#1}}

\def\sci#1{{ Science} {\bf#1}}

\def\rmp#1{{ Rev. \ Mod. \ Phys.} {\bf#1}}
\def\nat#1{{ Nature} {\bf#1}}

\def\njp#1{{ New. J. \ Phys.} {\bf#1}}

\begin{document}

\title{{
Adiabatic Quantum Simulation Using Trotterization}}

\author{Yifan Sun$^{1,2,3}$, Jun-Yi Zhang$^{1}$, Mark S. Byrd$^{4}$, Lian-Ao Wu$^{2,3}$}\thanks{Author to whom any correspondence should be addressed. Email address: lianao.wu@ehu.es}

\affiliation{$^{1}$State Key Laboratory of Magnetic Resonance and Atomic and Molecular Physics, Wuhan Institute of Physics and Mathematics, Chinese Academy of Sciences, Wuhan 430071, China \\ $^{2}$Department of Theoretical Physics and History of Science, The Basque Country University (EHU/UPV), PO Box 644, 48080 Bilbao, Spain \\ $^{3}$Ikerbasque, Basque Foundation for Science, 48011 Bilbao, Spain \\ $^{4}$Physics Department and Computer Science Department,
Southern Illinois University, Carbondale, Illinois 62901-4401}

\date{\today}

\begin{abstract}
As first proposed for the adiabatic quantum information processing by Wu, Byrd and Lidar [\prl{89}, 057904 (2002)], the Trotterization technique is a very useful tool for universal quantum computing, and in particular, the adiabatic quantum simulation of quantum systems.
Given a boson Hamiltonian involving arbitrary bilinear interactions, we propose a {\em static} version of this technique to perform an optical simulation that would enable the identification of the ground state of the Hamiltonian.  By this method, the dynamical process of the adiabatic evolution is mapped to a static linear optical array which avoids the errors caused by dynamical fluctuations. We examine the cost of the physical implementation of the Trotterization, i.e. the number of discrete steps required for a given accuracy. 
Two conclusions are drawn.  One is that number of required steps grows much more slowly than system size if the number of the non-zero matrix elements of Hamiltonian is not too large. The second is that the fluctuation of the parameters of optical elements does not affect the first conclusion.  This implies that the method is robust against errors. 
\end{abstract}


\maketitle

The reason for simulating a quantum system using another quantum system is to obtain information of an uncontrollable system from a controllable one which is similar to the former. It has attracted a lot of attention ever since proposed by Richard P. Feynman \cite{Feynman1982}, and developed by Seth Lloyd \cite{Lloyd1996}. Recent studies \cite{Abrams1997,Farhi2001,Wu2002,Martonak2004,WuByrd,Georgescu2014,Rosales2016,Rosales2018,Wang2016}
show that quantum simulation can provide alternative approaches to finding solutions by encoding them to the ground state of a Hamiltonian. Some of the simulation strategies have been proven to be capable of dealing with classically intractable problems, for example NP-complete problems \cite{Farhi2001,Wang2016}.

One major obstacle to realizing the quantum simulation of a particular system is the difficulties in preparing the ground state of a Hamiltonian.  In a number of quantum systems, it is relatively easy to find the ground state of some particular Hamiltonian, but very difficult to find the one required to solve a specific problem about which we are concerned.
A great deal of effort has been expended developing the strategies and technologies for ground state preparation, both experimentally and theoretically \cite{Teufel2011,Cormick2013,Jing2016}. Among those preparation strategies, adiabatic evolution has the greatest generality. In principle, if one prepares the ground state of some Hamiltonian, one can then  obtain the ground state of a target Hamiltonian by starting with the ground state that one can prepare and slowly evolving the system from the prepared Hamiltonian to the desired one. Such a scheme is guaranteed by adiabatic theorem and now termed {\it adiabatic quantum computing} (AQC) \cite{Albash2018}. AQC has been verified by a group of experiments \cite{Steffen2003,Johnson2011,Boixo2014,Barends2016,Wang2018}. 
It is considered a competitive candidate for universal quantum computing. In the implementation of AQC, the crucial step is to adiabatically connect the problem Hamiltonian (whose ground state encodes the solution) with the initial, prepared  Hamiltonian. Fortunately, the Trotterizaion technique provides a way to achieve such connection. 
With this technique, one can decompose the total evolution into short-time operations during which the system Hamiltonian is approximately time-independent for each step. The dynamical control of the system can be implemented by a sequence of such an operation.
This dramatically lowers the difficulty of realizing AQC. In general, the whole Trotterized-AQC (TAQC) protocol can be described as follows \cite{Wu2002}. (i) Prepare the ground state $|\psi_0\rangle$ of Hamiltonian $H_0$. 
(ii) Find the problem Hamiltonian $H_p$ whose ground state encodes the solution. (iii) Set the total Hamiltonian $H(t)=f(t)H_0+g(t)H_p$ with slowing-varying control functions $f(t)$ and $g(t)$, e.g.,  $f(t)=1-t/T$ and $g(t)=t/T$ where $t$ is the time and $T$ is the period for the entire evolution.  Then decompose the evolution operator into a sequence of steps using the Trotter-Suzuki formula, which is the key ingredient and given by   
\begin{eqnarray}\label{Evolve}
        U(T)&:=&\mathcal{T}\exp[-i\int_0^TH(t)dt] \nonumber \\
        &\approx&\prod_{a=0}^{k-1}\exp[-iH(a\tau)\tau].
\end{eqnarray}
$U(T)$ is the evolution operator from $0$ to $T$, $k$ is a large integer so that $\tau=T/k$ is a small time segment, and $\mathcal{T}$ is time ordering operator. (iv) Finally, obtain the solution by measuring the state $|\psi_f\rangle$ which is the simulation of $|\psi_{ad}\rangle=U(T)|\psi_0\rangle$ using $U(T)$ implemented according to Eq.~(\ref{Evolve}).  For operators $A$ and $B$ and a sufficiently small $\delta$, the Trotter-Suzuki formula implies $e^{\delta(A+B)}\approx e^{\delta A}e^{\delta B}+O(\delta^2)$. It was introduced for the simulation of complex time-independent Hamiltonians in Ref.~\cite{Lloyd1996}.  The application of the formula to an adiabatic strategy involving time-dependent Hamiltonian in TAQC protocol described above, was first proposed in Ref.~\cite{Wu2002} and experimentally verified by reference \cite{Barends2016}.

Here, we propose an optical implementation of TAQC. Linear optics provides a reasonably good system for quantum information processing. A logical qubit can be encoded in the polarization, frequency, spatial modes or other degrees of freedom of a photon which can be preserved for a relatively long time and is controllable  \cite{DiVincenzo2000,Knill2001,Kok2007,Sun2014,Andersen2015,Masada2015,Carolan2015,Tang2018}.   Just as important for our purposes, the operations of the system are static so that the dynamics are discretized.  We consider the problem of diagonalizing a matrix which is classically classified as NP-hard.  In order to simulate a many-body system, we propose a method for reaching the ground state of a boson Hamiltonian with arbitrary bilinear interactions. We analyze the dependence of the implementation cost, given by the Trotter Number (parameter $k$ in the decomposition (\ref{Evolve})), on the system size. We also study the effects of fluctuations of the parameters of the simulation by using a Randomized Trotter formula (RTF) \cite{Wang2016}. The definition of RTF is 
\begin{equation}\label{RTF}
    \begin{split}
        U(T)\approx\prod_{a=0}^{k-1}\exp[-iH(a\tau)\tau_a],
    \end{split}
\end{equation}
where $\tau_a=\tau(1+g_a)$ and $g_a$ is a random number. When $g_a$ is deleted, the decomposition (\ref{RTF}) reduces to the standard one (\ref{Evolve}).  In our case, the fluctuation of $\tau$ corresponds to the imperfections of experimental optical elements. We show numerically that such error will add little extra cost to the simulation for a given accuracy. 

We consider a very general model with 
\begin{equation}\label{Hami}
    \begin{split}
        H_0=\sum_s\epsilon_s{b_s^{\dagger}}b_s,~H_p=\sum_l\varepsilon_lb_l^{\dagger}b_l+\sum_{m\neq n}J_{mn}b_m^{\dagger}b_n.
    \end{split}
\end{equation}
where $b_i^{\dagger}(b_i)$ is a creation (annihilation) operator of the $i$th bosonic mode with commutators $[b_i,b_j^{\dagger}]={\delta}_{ij}$, $[b_i^{\dagger},b_j^{\dagger}]=[b_i,b_j]=0$ and $J_{mn}$ is the magnitude of interaction between the $m$th mode and $n$th mode. In the one-photon subspace, the Hamiltonian (\ref{Hami}) can be represented by a matrix which has no additional constraints other than being Hermitian. So the process of finding its ground state is equivalent to diagonalizing a general Hermitian matrix. In our proposal, the bosonic modes are mapped to the spatial modes of photons. Hence, $b_i^{\dagger}$ corresponds to a photon propagating along an optical path labelled by $i$, and $b_i$ corresponds to the absence of the photon from the path. To implement a TAQC, one must design a physical realization of the adiabatic evolution. We now discuss the details of such a realization. First, applying the decomposition (\ref{Evolve}) to the evolution of a Hamiltonian (\ref{Hami}), we have 
\begin{equation}\label{E0}
    \prod_{a=0}^{k-1}e^{(1-a/k)\tau\sum_s\epsilon_s{b_s^{\dagger}}b_s+(a/k)\tau(\sum_l\varepsilon_lb_l^{\dagger}b_l+\sum_{m\neq n}J_{mn}b_m^{\dagger}b_n)}.
\end{equation}
We can utilize the Hermiticity of $J$, $J_{mn}=J_{nm}^{*}$, so that
\begin{equation}
    \begin{split}
        &\sum_{m\neq n}J_{mn}b_m^{\dagger}b_n =\sum_{m<n}(J_{mn}b_m^{\dagger}b_n+J_{mn}^*b_mb_n^{\dagger})\\
        & =\sum_{m<n}[\textrm{Re} J_{mn}(b_m^{\dagger}b_n+b_mb_n^{\dagger})+i\textrm{Im} J_{mn}(b_m^{\dagger}b_n-b_mb_n^{\dagger})],
    \end{split}
\end{equation}
where $\textrm{Re}J_{mn}$ ($\textrm{Im}J_{mn}$) is the real (imaginary) part of $J_{mn}$. Given the commutators and Trotter-Suzuki formula, every multiplier of expression (\ref{E0}) can be separated into three exponential operators, and each one can be further decomposed as
\begin{equation}\label{D1}
    \begin{split}
        e^{-i(1-a/k)\tau\sum_s\epsilon_s{b_s^{\dagger}}b_s}&=\prod_se^{-i(1-a/k)\tau\epsilon_s{b_s^{\dagger}}b_s},\\
        e^{-i(a/k)\tau\sum_l\varepsilon_lb_l^{\dagger}b_l}&=\prod_le^{-i(a/k)\tau\varepsilon_lb_l^{\dagger}b_l},\\
    \end{split}
\end{equation}
and
\begin{equation}\label{D2}
    \begin{split}
        e^{-i(a/k)\tau\sum_{m{\neq}n}J_{mn}b_m^{\dagger}b_n}&\approx\prod_{m<n}\left[e^{(a/k)\tau\textrm{Im} J_{mn}(b_m^{\dagger}b_n-b_mb_n^{\dagger})}\right.\\
        &\left.\times e^{-i(a/k)\tau\textrm{Re} J_{mn}(b_m^{\dagger}b_n+b_mb_n^{\dagger})}\right].
    \end{split}
\end{equation}

\begin{figure}[htbp]
\centering
\includegraphics[width=3.1in]{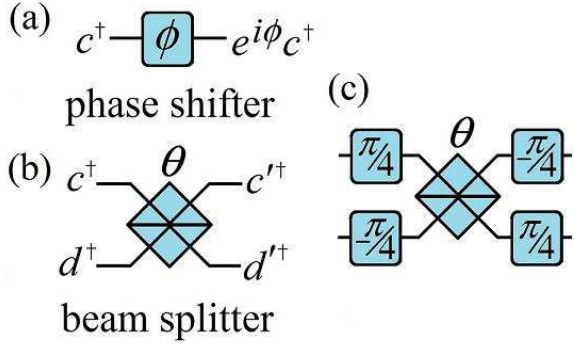}
\caption{Optical elements and their combinations used for simulation. (a) Phase shifter (b) Beam splitter and (c) combination for the simulation of the real part of the interaction. (Details are in the main text.) Output modes $c'^{\dagger}$ and $d'^{\dagger}$ in (b) are defined by $c'^{\dagger}=c^{\dagger}\cos{\theta}+d^{\dagger}\sin{\theta}$, $d'^{\dagger}=-c^{\dagger}\sin{\theta}+d^{\dagger}\cos{\theta}$.
}\label{DF1}
\end{figure}

\begin{figure}[htbp]
\centering
\includegraphics[width=3.4in]{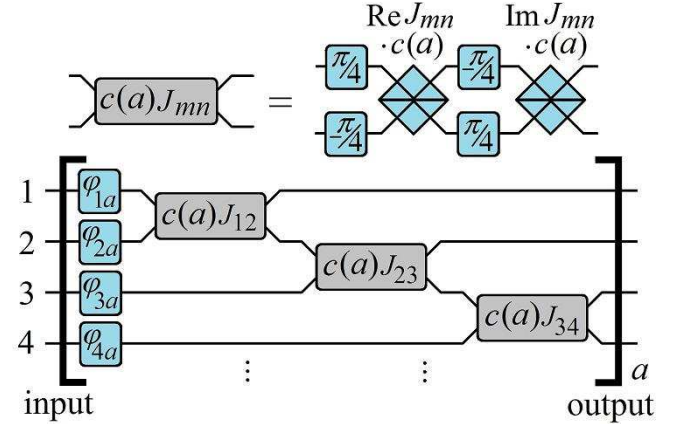}
\caption{Sketch of the whole simulation of the adiabatic evolution (for only nearest-neighbor interactions). The parameters of each element are shown in the figure. Square brackets labeled by $a$ mark out the unit cell which periodically repeats along the propagation direction of photons (from input to output) with $a=0,\cdots,k-1$. Function $c(a)=\tau a/k$. Phase function of the PS $\varphi_{na}=(1-a/k)\tau_a\epsilon_n+(a/k)\tau_a\varepsilon_n$.}\label{DF2}
\end{figure}

Next, we demonstrate how to simulate these operators using an array of optical devices.  However, we note that it is possible to implement the same set of elements using a photonic chip \cite{Tang2018}.  Our proposal could be considered a prototype for such a chip. We primarily use two common linear elements, phase shifters (PSs) and beam splitters (BSs), shown by Figs. \ref{DF1}(a) and \ref{DF1}(b). The mathematical descriptions of PS and BS are $U_{\rm{ps}}(\phi)=e^{-i{\phi}c^{\dagger}c}$ and $U_{\rm{bs}}(\theta)=e^{\theta(c^{\dagger}d-cd^{\dagger})}$.  (See for example \cite{Nielsen2000}.)  $c^{\dagger}$ and $d^{\dagger}$ are two different spatial modes.  $\phi$ is the phase shifted by a PS and $\theta$ defines the reflection (transmission) rate of a BS through $\cos\theta$ ($\sin\theta$).  $U_{\rm{ps}}(\phi)$ and $U_{\rm{bs}}(\theta)$ perfectly match the form of equations (\ref{D1}) and (\ref{D2}).  The factors of the forms $e^{-i(1-a/k)\tau\epsilon_s{b_s^{\dagger}}b_s}$ and $e^{-i(a/k)\tau\varepsilon_lb_l^{\dagger}b_l}$ can be implemented by two PSs, $U_{\rm{ps}}^s((1-a/k)\tau\epsilon_s)$ and $U_{\rm{ps}}^l((a/k)\tau\varepsilon_l)$. Superscripts $s$ and $l$ denote the optical modes.  Factor $e^{(a/k)\tau\textrm{Im} J_{mn}(b_m^{\dagger}b_n-b_mb_n^{\dagger})}$ can be implemented by one BS, $U_{\rm{bs}}^{mn}((a/k)\tau\textrm{Im} J_{mn})$, where $m$ and $n$ denote optical modes.  The factor $e^{-i(a/k)\tau\textrm{Re} J_{mn}(b_m^{\dagger}b_n+b_mb_n^{\dagger})}$ can be implemented by a combination of four PSs and a BS, $U^m_{\rm{bs}}(\frac{-\pi}{4})U^n_{\rm{bs}}(\frac{\pi}{4})U^{mn}_{\rm{bs}}((a/k)\tau\textrm{Re} J_{mn})U^m_{\rm{bs}}(\frac{\pi}{4})U^n_{\rm{bs}}(\frac{-\pi}{4})$. This can be seen by using the relation $e^{-\frac{\pi}{4}Z}Ye^{\frac{\pi}{4}Z}=X$ and the connection between the Lie group $SU(2)$ and boson operators. An illustration of the above combination is given by Fig. \ref{DF1}(c). The whole implementation of the simulation is described by Fig. \ref{DF2}.  For ease of illustration, Fig. {\ref{DF2}} only shows nearest-neighbour interactions. However, it is in principle possible to implement any type of bilinear interaction.

Our objective is the simulation of large quantum systems which are difficult to simulate using classical computers. However, when system size grows, more resources may be required to obtain the same level of simulation accuracy. Therefore, it is important to examine the variation of the resources with the system size.  We next investigate this resource dependence in terms of the number of required segments ($k$) when the number of bosonic modes ($N$) increases. 

We note that, as shown by the decomposition (\ref{Evolve}) and the expression (\ref{E0}), the accuracy of simulation increases when Trotter number $k$ grows. Also, the number of optical elements required to perform the simulation is proportional to $k$ (see Fig. \ref{DF2}).  Now consider the difference between the ideal adiabatic evolution and the Trotterized one as measured by $\Delta=1-|\langle\psi_{ad}|\psi_f\rangle|^2$.  The function $U_d(T)$, which is the discrete form of $U(T)$ obtained using a finite-difference Sch{\"o}dinger equation, is
\begin{equation}\label{UDT}
    \begin{split}
        U_d&(T)=1+(-i\tau)\sum_{r_1=0}^{k-1}H(r_1\tau)\\ 
        &+(-i\tau)^2\sum_{r_1=1}^{k-1}H(r_1\tau)\sum_{r_2=0}^{r_1-1}H(r_2\tau)+O(\tau^3).\\
    \end{split}
\end{equation}
Obviously, $U_d(T)\to U(T)$ when $k\to\infty$.  Also, consider the commutator of $H_0$ and $H_p$ which, to a large extent, describes the error when applying equations (\ref{D1}) and (\ref{D2}). By Taylor expansion, we can find the difference of $\prod_{a=0}^{k-1}e^{-iH(a\tau)\tau}$ and $\prod_{a=0}^{k-1}e^{-i(1-a/k)H_0\tau}e^{-i(a/k)H_p\tau}$. Then, the dominant factor of $\Delta$ can be calculated by adding up the leading terms in above expressions. More specifically, from $U(T)$ to $U_d(T)$, we have the leading error given by
\begin{equation}\label{DH1}
    D_1=\frac{H_p-H_0}{2}\tau.
\end{equation}
From $U_d(T)$ to $\prod_{a=0}^{k-1}e^{-iH(a\tau)\tau}$ (expression (\ref{E0})), the leading error is given by
\begin{equation}\label{DH2}
    D_2=-\frac{(-i\tau)^2}{2}\sum_{\alpha=0}^{k-1}H^2(\alpha\tau).
\end{equation}
From $\prod_{a=0}^{k-1}e^{-iH(a\tau)\tau}$ to $\prod_{a=0}^{k-1}e^{-i(1-a/k)H_0\tau}e^{-i(a/k)H_p\tau}$ which is, in principle, sufficient to describe the $k\sim N$ relation, the leading error is given by
\begin{equation}\label{DH3}
    D_3=-\sum_{\beta=0}^{k-1}\frac{(-i\tau)^2}{2}[H_p(\beta\tau),H_0(\beta\tau)].
\end{equation}
Then we have
\begin{equation}
    \begin{split}
       |\langle\psi_{ad}|\psi_f\rangle|^2&\approx|\langle\psi_0|U^{\dagger}(T)[U(T)-D]|\psi_0\rangle|^2\\
       &=|1-\langle\psi_0|U^{\dagger}(T,0)D|\psi_0\rangle|^2\\
       &\approx 1-2\mathrm{Re}\{\langle\psi_0|U^{\dagger}(T)D|\psi_0\rangle\}\\
       &=1-2{\rm Re}\{\langle\psi_{ad}|D|\psi_0\rangle\},
    \end{split}
\end{equation}
where $D=D_1+D_2+D_3$. In the second to last approximation, higher order terms are neglected. After some simplification, we pick out the leading terms and obtain
\begin{equation}
    \begin{split}
        \Delta&\approx\frac{T^2}{3k}{\rm Re}\{[\langle \psi_{ad}|H^2_0|\psi_0\rangle+\langle\psi_{ad}|H_pH_0|\psi_0\rangle\\
        &-\frac{3}{2T}\langle\psi_{ad}|(H_p-H_0)|\psi_0\rangle+\langle\psi_{ad}|H^2_p|\psi_0\rangle]\}\\
        &=\frac{T^2}{3k}{\rm Re}\{\langle\psi_{ad}|\psi_0\rangle\}[E_{0g}^2+E_{0g}E_{pg}\\
        &-\frac{3}{2T}(E_{pg}-E_{0g})+E^2_{pg}],
    \end{split}
\end{equation}
where $E_{pg}$ ($E_{0g}$) is the ground state energy of $H_p$ ($H_0$).  This  expression comes from the fact that $|\psi_{ad}\rangle$ ($|\psi_0\rangle$) is the ground state of $H_p$ ($H_0$). Because $H_0$ is diagonal, $E_{0g}$ is independent of $N$. In general, $E_{pg}$ is a function of $N$ determined by the structure of $H_p$. The real part of the overlap $\langle\psi_{ad}|\psi_0\rangle$ is bounded by one. So we can rewrite $\Delta$ in the following form:
\begin{equation}\label{FD}
    \Delta\approx\frac{T^2}{3k}(A+BE_{pg}(N)+CE_{pg}^2(N)),
\end{equation}
where $A$, $B$, $C$ are constants independent of $N$. We can conclude from equation (\ref{FD}) that for a given $\Delta$, the dependence of the Trotter number $k$ on system size $N$ is determined by the ground state energy of $H_p$ in the leading order approximation.  Thus determining the form of $H_p$ will enable the determination of the dependence of $E_g$ on $N$ and therefore the relation between $k$ and $N$ for a given $\Delta$.

Next, we consider the effect of the fluctuation of the optical elements on this  relation. The analysis process is basically the same as before, except that we replace $\tau$ by $\tau_a$ (given by RTF (\ref{RTF})). The fluctuation is modeled by zero-mean random number $g_a$. Then one has $\sum_{a=0}^{k-1}g_a\to 0$ when $k\to\infty$. This means that the $g_a$ included in the summations will converge to zero like $1/k$ and can be treated as a higher order term. Then $\tau_a$ shrinks to $\tau$ in the leading order approximation and the dominant error in such case has the same form as Equation (\ref{FD}). Therefore, the fluctuations introduced by $\tau_a$ does not contribute significantly to the dependence of $k$ on $N$. 
\begin{figure}[htbp]
\centering
\includegraphics[width=3.2in]{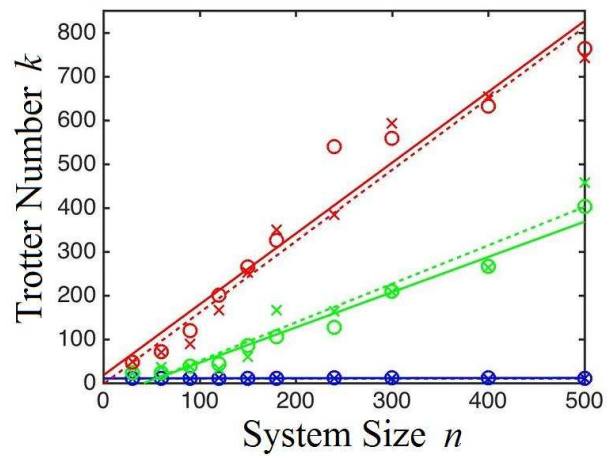}
\caption{The dependence of Trotter number $k$ on system size $N$ when the overlap $|\langle\psi_{ad}|\psi_f\rangle|^2$ is bigger than 0.9. Solid lines are fitted via data points marked by $\circ$ which are obtained by the ordinary Trotter decomposition. Dashed lines are fitted via data points marked by $\times$ which are obtained by RTF. The blue, green and red results correspond to the cases when $H_p$ is pentadiagonal, random sparse and fully random respectively.}\label{DF3}
\end{figure}

We next numerically evaluate this dependence for some particular cases.  We first let $H_0$ be a diagonal matrix whose entries are sorted, equal-spaced and from $0.5$ to $N-0.5$. The diagonal entries of $H_p$ are $H_0-0.5$. Before the description of the off-diagonal setup of $H_p$, we introduce the concept of the density of a matrix, which is defined as the number of nonzero matrix elements divided by the total number of matrix elements. We consider three off-diagonal examples of $H_p$. The first one only involves the nearest- and next-nearest-neighbour interaction, i.e. $H_p$ is a pentadiagonal matrix with density $(5N-6)/N^2$. The second off-diagonal part forms a sparse matrix with fixed density 0.5. The locations of non-zero entries are random. The third one is the case with full non-zero-off-diagonal entries which means the total density is 1. The values of the off-diagonal entries in all three types of $H_p$ randomly varies from 0 to 1. The simulation results are shown in Fig. \ref{DF3}. The value of $k$ is found by increasing from a small number till the moment when overlap $|\langle\psi_{ad}|\psi_f\rangle|^2$ is bigger than 0.9. $|\psi_f\rangle$ is obtained by numerical simulation of TAQC and $|\psi_{ad}\rangle$ is obtained by direct diagonalization of $H_p$. The value of $k$ of each point (marked by $\circ$ or $\times$) in Fig. \ref{DF3} is the average of eight $k$s, but the same $N$. The lines are fitted via linear regression. The solid lines are fitted by ordinary Totter decomposition data points marked by $\circ$ and the dashed lines are fitted by RTF data points marked by $\times$. The slope of the solid lines in Fig. \ref{DF3} are $2.10\times10^{-3}$, $0.80$, $1.62$ and that of the dashed lines are $1.46\times10^{-4}$, $0.88$, $1.63$, from the bottom to the top. These results can be explained by Equation (\ref{FD}). We simulate the change of $E_{pg}$ with $N$ in above three cases. For the first type of $H_p$ (bottom, blue), $E_{pg}$ only varies when $N$ increases, so $k$ is nearly constant. For the other two, $E_{pg}^2$ is found to be linear for both cases and the slope of the fitting lines are $9.20\times10^{-4}$ for the third and $5.09\times10^{-4}$ for the second, which is nearly half of the former. The linearity of $E_{pg}^2$ also indicates that $E_{pg}\sim\sqrt{N}$. This means that $E_{pg}^2$ is dominant, especially when $N$ is large. Therefore, from Fig. \ref{DF3}, we can see that the slope of the third (top, red) is approximately twice as much as the slope of the second (middle, green). The solid lines and the dashed lines are nearly coincident which supports our analysis of the fluctuations.

In conclusion, using the Trotterization technique, we proposed a scheme to adiabatically reach the ground state of a boson Hamiltonian with arbitrary bilinear interactions. The whole process is implemented by an linear optical design which is robust against errors caused by fluctuations in the accuracy of the individual elements. To the best of our knowledge, Trotterization is the only way a dynamical quantum process can be completely mapped to a static circuit. We also analyzed the dependence of implementation cost on the system size when the  simulation accuracy is approximately fixed.  Corresponding analytical and numerical results show that, the cost of the simulation, represented by Trotter number $k$, and system size $N$ is determined by the structure of problem Hamiltonian.  When the structure is rather simple, such as the case when the density of $H_p$ is not large, the cost will grow more slowly than system size. Moreover, we found that imperfect experimental conditions, modelled by parameter fluctuations, do not significantly affect the relation between $k$ and $N$ which means that the simulation is robust against errors.  

\acknowledgments
We acknowledge grant support from the Spanish MINECO/FEDER Grants FIS2015-69983-P, the Basque Government Grant IT986-16 and UPV/EHU UFI 11/55.

\end{document}